\documentclass[a4paper]{article}
\usepackage{geometry}
\usepackage{amsmath}
\newcommand{\bv}{\begin{vmatrix}}
\newcommand{\ev}{\end{vmatrix}}

\newcommand{\bea}{\begin{eqnarray*}}
\newcommand{\eea}{\end{eqnarray*}}
\newcommand{\bean}{\begin{eqnarray}}
\newcommand{\eean}{\end{eqnarray}}

\date{ }
\title{Naked Singularity as Accelerator for Spinning Particle}
\author{ Jincheng An\footnote{anjch@mail.bnu.edu.cn}   \\
Department of Physics, Beijing Normal University,\\
Beijing 100875, China}

\begin{document}
\maketitle
\begin{abstract}
 When two particles collide at the horizon of a black hole and one of them satisfies some critical conditions, the relative velocity between them can be arbitrarily large, thus the energy of the center-of-mass will reach infinity. Such a process is called BSW mechanism which can accelerate a particle to arbitrarily high energy. There are also some studies showing that a Kerr naked singularity can be more qualified  as a particle accelerator for arbitrarily high energy. Previous researchers mainly concentrate on geodesic motion of particles. In this paper, we will take spinning particles which won't move along a timelike geodesic and carry more parameters into our consideration. By employing the Mathisson-Papapetrou-Dixon equation, we will prove that for a spinning particle in hyper-extremal Reissner-Nordstrom or Kerr spacetime where exists a naked singularity at $r=0$, its Effective Potential  $V_{eff}=-\dot{r}^2$ must be able to reach zero within the interval $0 < r < M$,  thus an ingoing particle will be able to turn back and then collide with another ingoing particle at $r=M$. If the spacetime is ${\bf slightly}$ hyper-extremal, the energy of center of mass $E_{cm}$ will be arbitrarily high.
\end{abstract}
\section*{Introduction}
Ba\~nados, Silk, and West (BSW mechanism) firstly proved \cite{BSW} that an extremal Kerr black hole can be used as a particle accelerator in 2009, and the center-of-mass (CM) energy $E_{cm}$ of two test particles can be arbitrary high if the collision occurs near the extremal horizon and the energy $E$ and angular momentum $L$ of one particle satisfy a critical relation. Based on their pioneering discussion, the BSW mechanism has been extensively studied for multifarious black hole backgrounds\cite{272}-\cite{1025}. Since the event horizon is a one-way membrane which means that once an ingoing particle passed the horizon, it would never come out of the horizon again, hence the movement of particles is heavily restricted by the existence of horizon. So there are some researchers  concentrating on the BSW mechanism near various naked singularities \cite{naked}\cite{naked1}\cite{RNnaked}\cite{naked2}, since the disappearance of the event horizon will give the movement of particles more freedom. Therefore, without the restriction of event horizon, an infinite $E_{cm}$  can be more easily obtained near a Kerr singularity, hence the critical relation \cite{BSW}\cite{4D} which is a harsh requirement can be relaxed largely \cite{naked}.

Most of the previous work on BSW mechanism focus on the acceleration of point particles which are not spinning and move along timelike geodesics. However, a real-world particle is an extended body with self-interaction, thus, it is more reasonable to take spinning particles which cannot be regarded as point particles into our consideration. It has been proved in\cite{Wald1972} and \cite{s1}\cite{s2}\cite{s3} that the movement of a spinning particle deviates from a geodesic due to the gravitational  interaction. The orbits of spinning particles has been computed based on the Mathisson-Papapetrou-Dixon (MPD) equation \cite{Motoyuki}\cite{z1}\cite{z2}. The BSW mechanism for spinning particles in Kerr space and Kerr-Newman (KN) spacetime has been studied in \cite{myg} and \cite{yuxiao},where for an ultrahigh $E_{cm}$, a critical relation is also needed and the black hole must be extremal. In the meantime, \cite{myg} and \cite{yuxiao} also gave out a critical spinning angular momentum that can also lead to a divergent $E_{cm}$ without the critical relation needing to be satisfied, which is typical for spinning particle. However, as argued in \cite{ks} where the acceleration of spinning particle by extremal Kerr-Sen black hole is studied, the critical spin for extremal Kerr or KN black hole is unapplicable. \cite{ks} presented a critical spin for extremal Kerr-Sen black hole, which is accessible in a real world. Spinning particle collision in Schwarzchild black hole background was discussed in \cite{zs}.

In this paper, we will focus on the acceleration of spinning particles by Reissner-Nordstrom singularity and Kerr naked singularity,  and we will prove that an ingoing spinning particle released from infinity towards the naked singularities can certainly turn back after it pass $r = M $ where the original extremal event horizon is located before it reach the naked singularities. Therefore, the turning-back spinning particle is able to collide with another ingoing particle at $r=M$.  We will also prove that in such case, the $E_{cm} $ will be divergent if the spacetime is slightly hyper-extremal, i.e. the extremal Reissner-Nordstrom black hole is slightly over-charged, $Q=M+\epsilon$, or the extremal  Kerr black hole is over-spun, $a=M+\epsilon$, where $\epsilon$ is positive and can be arbitrarily small.

The structure of this paper is as following: To start with, in section \ref{rnp}, we will discuss the collision of non-spinning particles near the Reissner-Nordstrom naked singularity and give out the general analytic method we will use in this paper, we will also prove an infinite $E_{cm}$ can be reached if the collision happened between ingoing and an outgoing particles, further, we proved an ingoing particle will certainly turn back within the interval $0<r<M$, thus, there does exist possibility for the collision to occur at $r=M$. Next in section \ref{rns}, we will figure out the movement of spinning particles in a general static spherical spacetime and apply the analytic method in section \ref{rnp} to spinning particle, similarly, we also proved that it is  possible an ingoing particle will certainly turn back when $0<r<M$, hence, an ultrahigh-energy collision can happen. Finally, in section \ref{kerrs}, with the help of tetrads, the movement of spinning particles in Kerr spacetime will be computed, and  the $E_{cm}$ of collision at $r=M $ between ingoing and outgoing spinning particles will be proved to be arbitrarily high, the  possibility for such collision to happen was also confirmed to exist.
\section{Particles Collision near Reissner-Nordstrom Naked Singularity }\label{rnp}
Let's start with a general static spherical black hole which reads
\begin{eqnarray}\label{stc}
ds^2=-fdt^2+\frac{dr^2}{f}+r^2(d\theta^2+\sin{\theta}^2d\phi^2),
\end{eqnarray}
where $f$ is a function of $r$ and the event horizon is located at $f(r_H)=0$ ($r_H$ is positive). Now  we consider the 4-velocity
\bean
\mu^a=\dot{t}\left(\frac{\partial}{\partial t}\right)^a+\dot{r}\left(\frac{\partial}{\partial r}\right)^a+\dot{\theta}\left(\frac{\partial}{\partial \theta}\right)^a+\dot{\phi}\left(\frac{\partial}{\partial \phi}\right)^a
\eean
of point particle moving along a timelike geodesic. After fixing $\theta$ at $\frac{\pi}{2}$ (thus, $\dot{\theta}=0
$) and using the $3$ equations
\begin{eqnarray}\label{n1}
&&g_{ab}\mu^a\mu^b=-1,\label{n1}\\
&&g_{ab}\mu^a\xi^b=-E,\label{en}\\
&&g_{ab}\mu^a\phi^b=L,\label{an}
\end{eqnarray}
where
\bean
\xi^a=\left(\frac{\partial}{\partial t}\right)^a \quad \text{and}\quad \phi^a=\left(\frac{\partial}{\partial \phi}\right)^a
\eean
 are the Killing  vectors associated with 2 conserved quantities, the energy $E$ and the angular momentum $L$, we can get that
\begin{eqnarray}
&&\dot{t}=\frac{E}{f},\label{td}\\
&&\dot{r}=\pm\frac{\sqrt{E^2r^2-f(L^2+r^2)}}{r},\label{rd}\\
&&\dot{\phi}=\frac{L}{r^2}.\label{fd}
\end{eqnarray}
As defined in \cite{BSW}, the energy of the center of mass is
\begin{eqnarray}
E_{cm}=\sqrt{2}m\sqrt{1-g_{ab}\mu_1^a\mu_2^b}.
\end{eqnarray}
For convenience, we only need to focus on the changeable part of $E_{cm}$, which is defined \cite{4D} as
\begin{eqnarray}\label{eff}
E_{eff}=-g_{ab}\mu_1^a\mu_2^b,
\end{eqnarray}
which, physically speaking, describes the relative velocity between the two colliding particles. Manifestly, when the relative velocity between particles becomes infinite, the $E_{cm}$ can reach infinity.

Let's consider the collision between two point particles near the horizon $f(r_H)=0$. After taking \eqref{td}\eqref{rd}and \eqref{fd} into \eqref{eff}, we get that
\begin{eqnarray}\label{efe}
E_{eff}&=&\frac{E_1E_2r^2-fL_1L_2\pm\sqrt{E_1^2r^2-(L_1^2+r^2)f}\sqrt{E_2^2r^2-(L_2^2+r^2)f}}{fr^2}\nonumber\\
&=&\frac{E_1E_2r^2-fL_1L_2\pm E_1E_2r^2\sqrt{1-\frac{(L_1^2+r^2)f}{E_1^2r^2}}\sqrt{1-\frac{(L_2^2+r^2)f}{E_2^2r^2}}}{fr^2}\nonumber\\
&=&\frac{E_1E_2r^2-fL_1L_2\pm E_1E_2r^2\left(1-\frac{(L_1^2+r^2)f}{2E_1^2r^2}\right)\left(1-\frac{(L_2^2+r^2)f}{2E_2^2r^2}\right)}{fr^2}\nonumber\\
&=&\frac{E_1E_2\pm E_1E_2}{f}+\alpha_{0\pm}+\alpha_{1\pm}f,
\end{eqnarray}
where $\alpha_{1\pm}$ and $\alpha_{2\pm}$ are both manifestly finite as long as $r_H>0$. The $Plus $ $sign$ denotes the collision between an outgoing particle and an ingoing particle while the $Minus $  $sign$  denotes the one between  two  both-ingoing or both-outgoing particles. Meanwhile, it is worth noting that
\begin{eqnarray}
\alpha_{0-}=\frac{E_1^2(L_2^2+r^2)E_2^2(L_1^2+r^2)-2E_1E_2L_1L_2}{2E_1E_2r^2},
\end{eqnarray}
when $E_1=E_2$, $L_1=L_2$,  we have $\alpha_{0-}=1$, which is consistent with \eqref{n1}\eqref{eff}.
So we can draw the conclusion that near the horizon, only the collision between an ingoing particle and an outgoing particle will lead to a divergence of $E_{eff}$, namely a infinity of $E_{cm}$.

Next, we define the effective potential as below
\begin{eqnarray}
V_{eff}=-\dot{r}^2.
\end{eqnarray}
 $V_{eff}<0$ means that a particle is attracted by the black hole to fall into it, while  $V_{eff}>0$ means that a particle is repulsed by the black hole to leave far away from it. It can be easily checked that for the Schwarzchild black hole ( $f=1-\frac{2M}{r}$ ) and Reissner-Nordstrom black hole ( $f=1-\frac{2M}{r}+\frac{Q^2}{r^2}$ ) , $V_{eff}<0$ outside the outer horizon, which means that if we release a point particle from infinity, the particle will keep going towards the black hole and it will not turn back at some point $r>r_H$. So we cannot get a infinite $E_{cm}$ from the collision near the horizon between two particles released from infinity. Thus we turn to  use naked  singularity as particle accelerator. We want to find a naked singularity near which a ingoing particle can turn back and then collide with another ingoing particle, thus we may be more likely to get a infinite $E_{cm}$.

 At first, we discuss the simplest naked singularity, the  Reissner-Nordstrom singularity. As mentioned above, for a Reissner-Nordstrom black hole
 \begin{eqnarray}\label{rnf}
 f(r)=1-\frac{2M}{r}+\frac{Q^2}{r^2}=\frac{r^2-2Mr+Q^2}{r^2},
\end{eqnarray}
its horizon is located at
\bean
r_{\pm} = M\pm\sqrt{M^2-Q^2}.
\eean
Only when $Q\leq M$ can the horizon exist, and as long as $ Q $ become  slightly larger than $M$, the horizon at $r = M$ will disappear abruptly.  We rewrite $Q$ as
\begin{eqnarray}
Q=M+\epsilon ,
\end{eqnarray}
where $\epsilon$ is a positive infinitesimal value, thus near  $r=M$ where the original extremal horizon is located, $f(r)$ can be re-expressed as
\begin{eqnarray}
f(r)=\frac{(r-M)^2+2M \epsilon}{r^2}.
\end{eqnarray}
We require that the collision will take place at $r=M$ , then we get that
\begin{equation}
E_{eff}\mid_{r=M}=M\frac{E_1E_2\pm E_1E_2}{2\epsilon}+\text{finite terms}.
\end{equation}

Next, we need to figure out whether an ingoing particle can turn back at some point $r_T$ which satisfied $0 < r_T < M $ , i.e. the equation $ V_{eff}(r)=0 $ has a root in the interval $0 < r < M $. After taking \eqref{rnf} into \eqref{rd}, we get $V_{eff}$ as
\begin{eqnarray}\label{vrn}
V_{eff}(r)=\frac{(1-E^2)r^4-2Mr^3+(L^2+Q^2)r^2-2L^2Mr+L^2Q^2}{r^4}.
\end{eqnarray}
To guarantee that a particle at infinity can go inwardly, $V_{eff}\mid_{r\rightarrow\infty}$ must be negative, thus, we need to set
\begin{eqnarray}
 E \geq 1 \nonumber.
\end{eqnarray}
 From \eqref{vrn}, we can see that
\begin{eqnarray}
V_{eff}\mid_{r\rightarrow0}=+\infty >0\nonumber ,
\end{eqnarray}
thus to make sure that  $ V_{eff}(r)=0 $ has a root in the interval $0 < r < M $ , we only need to request $V_{eff}$ to be negative at $r = M$.

Without losing generality and for convenience,  in the rest of this paper, we take
\bean
M=1,
\eean
  thus, $Q > 1 $, then we get that
\begin{eqnarray}
V_{eff}\mid_{r=1}=L^2(Q^2-1)+Q^2-E^2-1<0.
\end{eqnarray}
If we take $Q=1+\epsilon $ into it and let $\epsilon\rightarrow0$, namely $Q\rightarrow1$, we will find that
\begin{eqnarray}
\left[L^2(Q^2-1)+Q^2-E^2-1\right]\mid_{Q\rightarrow1}=-E^2<0.
\end{eqnarray}
then we have
\bea
\left(V_{eff}\mid_{r=0}\right)\left(V_{eff}\mid_{r=1}\right)<0,
\eea
there definitely exists a root for $V_{eff}=0$ within $0<r<1$. So we can assert
 that an ingoing point particle can turn back in the interval $0 < r < 1 $, hence it is possible for the collision between ingoing and outgoing particles to happen at $r=1$,  so we can get a infinity of $E_{cm}$ from such a collision.
\section{Collision of Spinning Particle near Reissner-Nordstrom Naked Singularity}\label{rns}
Inspire by the discussion above, let's explore how the $E_{cm}$ behaves when two spinning particles collide at $r=1$ in a over-charged Reissner-Nordstrom spacetime admitting a naked singularity.  To describe the movement of spinning particles, we will employ the Mathisson-Papapetrou-Dixon equation \cite{Wald1972}\cite{myg} :
\begin{eqnarray}
&&\frac{DP^a}{D\tau}=-\frac{1}{2}R^a _{bcd}v^bS^{cd},\\
&&\frac{DS^{ab}}{D\tau}=P^av^b-P^bv^a,
\end{eqnarray}
where
\begin{eqnarray}
v^a=\left(\frac{\partial}{\partial\tau}\right)^a
\end{eqnarray}
is the tangent to the center of mass world line, and $P^a$ is the 4-momentum of the spinning particle satisfying that
\begin{eqnarray}\label{n1s}
-m^2=P^aP_a,
\end{eqnarray}
$\frac{D}{D\tau}$ is the covariant derivative along the world line, $S^{ab}$ is the particle's spin tensor which is apparently antisymmetry
\bean
S^{ab}=-S^{ba},
\eean
 and has  property \cite{myg}
\begin{eqnarray}\label{ans}
J^2=\frac{1}{2}S^{ab}S_{ab}.
\end{eqnarray}
There also exists the relation between $S^{ab}$ and $P^a$ as below
\begin{eqnarray}\label{sp}
S^{ab}P_b=0.
\end{eqnarray}
For a spinning particle, the conserved quantities associated with the Killing  vectors in its movement can be expressed as
\begin{eqnarray}\label{scq}
C_{\xi}=P^a\xi_{a}-\frac{1}{2}S^{ab}\bigtriangledown_a\xi_b.
\end{eqnarray}
Let's consider the movement in the plane $\theta=\frac{\pi}{2}$, thus the 4-momentum can be assumed to be
\bean
P^a=P^t\left(\frac{\partial}{\partial t}\right)^a+P^r\left(\frac{\partial}{\partial r}\right)^a+P^{\phi}\left(\frac{\partial}{\partial \phi}\right)^a,
\eean
in the meantime, the non-zero components of the spin tensor $S^{ab}$ are assumed to be
\begin{eqnarray}
 S^{tr}=-S^{rt} ,\quad S^{t\phi}=-S^{\phi t},\quad S^{r\phi}=-S^{\phi r},
\end{eqnarray}
  meaning the direction of the particle's spin is normal to the equatorial surface $\theta=\frac{\pi}{2}$. In a general static spherical spacetime \eqref{stc}, equations \eqref{n1s} and \eqref{ans} can be explicitly expressed as
\begin{eqnarray}\label{n1se}
-m^2=-(P^t)^2f+\frac{(P^r)^2}{f}+r^2(P^{\phi})^2,
\end{eqnarray}
\begin{eqnarray}\label{anse}
J^2=-r^2(S^{t\phi})^2f+\frac{(S^{r\phi})^2}{f}-(S^{tr})^2.
\end{eqnarray}
Two Killing vectors $(\frac{\partial}{\partial t})^a$ and $(\frac{\partial}{\partial\phi})^a$  are associated two conserved quantities as below
\begin{eqnarray}\label{sen}
-E=-fP^t+\frac{1}{2}S^{tr}f^\prime,
\end{eqnarray}
\begin{eqnarray}\label{san}
L=P^{\phi}r^2+rS^{r\phi},
\end{eqnarray}
obviously, when $S^{ab}$ vanishes, the two equation will be reduced to \eqref{en} and \eqref{an}.

The relation \eqref{sp} appears explicitly as
\begin{eqnarray}
&&-P^{\phi}r^2S^{t\phi}-\frac{P^rS^{tr}}{f}=0,\label{spt}\\
&&-P^{\phi}r^2S^{r\phi}-P^tS^{tr}f=0,\label{spr}\\
&&-P^tS^{t\phi}f+\frac{P^rS^{r\phi}}{f}=0.\label{spf}
\end{eqnarray}
With equations \eqref{spt} and \eqref{spr}, we can get the relation that
\begin{eqnarray}
\frac{r^4(P^{\phi})^2(S^{r\phi})^2}{f}-f(P^{\phi})^2(S^{t\phi})^2r^4=f(P^t)^2(S^{tr})^2-\frac{(P^r)^2(S^{tr})^2}{f},
\end{eqnarray}
then combined with \eqref{n1se} and \eqref{anse}, it can be found that
\begin{eqnarray}\label{r=m}
r^2(P^\phi)^2J^2=m^2(S^{tr})^2.
\end{eqnarray}
Using equation \eqref{sen}\eqref{san} and \eqref{spr}, a  new relation can be drawn as below
\begin{eqnarray}\label{le=rfp}
LrP^\phi+ES^{tr}=r^3(P^\phi)^2-\frac{1}{2}f^\prime(S^{tr})^2.
\end{eqnarray}
Based on equations\eqref{sen}\eqref{r=m} and \eqref{le=rfp}, we can get that
\begin{eqnarray}
&&P^t=\frac{2m^2rE\pm mLJf^\prime}{(2m^2r-J^2f^\prime)f},\label{pt}\\
&&P^\phi=\frac{2m(Lm\pm EJ)}{r(2m^2r-J^2f^\prime)}.\label{pf}
\end{eqnarray}
Also, it can be checked that with $J=0$, $P^t$ and $P^\phi$ is just \eqref{td} and \eqref{rd} .
Then with \eqref{n1se}, we can figure out
\begin{eqnarray}\label{pr}
P^r=\pm\sqrt{f}\sqrt{(P^t)^2f-r^2(P^\phi)^2-m^2}.
\end{eqnarray}
As required by a future-directed particle, we need to set $P^t > 0 $ outside the horizon ($f > 0$), thus
\begin{eqnarray}\label{tdp}
\frac{2m^2rE\pm mLJf^\prime}{2m^2r-J^2f^\prime} > 0.
\end{eqnarray}
Now, without losing generality, we choose the $Plus$ in \eqref{pt}\eqref{pf}, and take them into \eqref{pr}, then $P^r$ becomes
\begin{eqnarray}\label{pre}
P^r&&=\pm\sqrt{\left(\frac{2Em^2r+JLmf^\prime}{2m^2r-J^2f^\prime}\right)^2-\left[m^2+\frac{4m^2(EJ+Lm)^2}{(2m^2r-J^2f^\prime)^2}\right]f}\nonumber\\
&&=\pm\left(\frac{2Em^2r+JLmf^\prime}{2m^2r-J^2f^\prime}\right)\sqrt{1-Wf},
\end{eqnarray}
where
\begin{eqnarray}
W=\frac{m^2+\frac{4m^2(EJ+Lm)^2}{(2m^2r-J^2f^\prime)^2}}{(\frac{2Em^2r+JLmf^\prime}{2m^2r-J^2f^\prime})^2},
\end{eqnarray}
for the convenience of expression, which apparently is finite.

The energy of center-of-mass can be written as
\begin{eqnarray}\label{ecm}
E_{cm}&&=\sqrt{-g_{ab}(P_1^a+P_2^a)(P_1^b+P_2^b)}\nonumber\\
&&=\sqrt{2m^2-2g_{ab}P_1^aP_2^b}.
\end{eqnarray}
Here we also can define
\begin{eqnarray}\label{epp}
E_{eff}=-g_{ab}P_1^aP_2^b,
\end{eqnarray}
which is proportional to the relative velocity between the two spinning particles.

So with \eqref{pt}\eqref{pf}and\eqref{pre}, the $E_{eff}$ can be rewritten as
\begin{eqnarray}
E_{eff}&=&\frac{1}{f(2m^2r-J_1^2f^\prime)(2m^2r-J_2^2f^\prime)}[-4m^2f(E_1J_1+L_1m)(E_2J_2+L_2m)\nonumber\\
&&\pm m^2(2E_1mr+J_1L_1f^\prime)(2E_2mr+J_2L_2f^\prime)\sqrt{1-W_1f}\sqrt{1-W_2f}\nonumber\\
&&+m^2(2E_1mr+J_1L_1f^\prime)(2E_2mr+J_2L_2f^\prime)],
\end{eqnarray}
Restricted by \eqref{tdp}, we know that  $2m^2r-J^2f^\prime\neq0$, thus the denominator of $E_{eff}$ will not vanish outside horizon. Near the horizon, we can use the relation $\sqrt{1-Wf}=1-\frac{Wf}{2}$ to rewrite $E_{eff}$ as
\begin{eqnarray}\label{effs}
E_{eff}&=&\frac{-4m^2f(E_1J_1+L_1m)(E_2J_2+L_2m)+m^2E_1^\prime E_2^\prime\left[1\pm \left(1-\frac{W_1f}{2}\right)\left(1-\frac{W_2f}{2}\right)\right]}{f(2m^2r-J_1^2f^\prime)(2m^2r-J_2^2f^\prime)}\nonumber\\
&=&\frac{m^2E_1^\prime E_2^\prime\pm m^2E_1^\prime E_2^\prime}{f(2m^2r-J_1^2f^\prime)(2m^2r-J_2^2f^\prime)}+\beta_{0\pm}+\beta_{1\pm}f,
\end{eqnarray}
where we have denoted that
\begin{eqnarray}
E^\prime=2Emr+JLf^\prime,
\end{eqnarray}
and $\beta_{0\pm}$, $\beta_{1\pm}$ are both finite. Similarly, when $E_1=E_2$, $L_1=L_2$, $J_1=J_2$, it can be checked that $\beta_{0-}=m^2$, which is requested by \eqref{n1s} and \eqref{epp}.

Since  the form of \eqref{effs} is very similar to that of \eqref{efe}, we can use the same method to discuss the collision of spinning particles near Reissner-Nordstrom naked singularity.
Similarly, we can define the  Effective Potential  as
\begin{eqnarray}
V_{eff}=-(P^r)^2,
\end{eqnarray}
with \eqref{pt}\eqref{pf}and\eqref{pr}, its explicitly reads
\begin{eqnarray}
V_{eff}&=&\frac{1}{(f^{\prime}J^2-2m^2r)^2}[4(EJm)^2f+J^4(mf^\prime)^2f-(f^\prime fJLm)^2+8EJLfm^3 \nonumber\\
&&+4fL^2m^4-4Ef^\prime JLm^3r-4ff^\prime J^2m^4r-4E^2m^4r^2+rfm^6r^2].
\end{eqnarray}
In Reissner-Nordstrom spacetime,
\begin{eqnarray}
f^\prime(r)=\frac{2(r-Q^2)}{r^3},
\end{eqnarray}
thus , we get
\begin{eqnarray}
V_{eff}=\frac{J^4m^2Q^6+\sum_{i=1}^{9}\sigma_i r^i+m^4(E^2-m^2)r^{10}}{r^2(J^2Q^2-J^2r+m^2r^4)^2},
\end{eqnarray}
where $\sigma_i=\sigma_i(J,L,Q,E,m)$ are all finite. From above, we can apparently see that
\begin{eqnarray}\label{v0}
V_{eff}\mid_{r\rightarrow0}=+\infty .
\end{eqnarray}
By the same way, we take $Q=1+\epsilon$, then will gain that
\begin{eqnarray}
f^\prime \mid_{r=1}=-4\epsilon.
\end{eqnarray}
Combining it with $f(1)=2\epsilon$, we get
\begin{eqnarray}
V_{eff}\mid_{r=1}=\frac{-E^2m^4+\sum_{i=1}^{3}\lambda_i \epsilon^i}{(m^2+2J^2\epsilon)^2},
\end{eqnarray}
where $\lambda_i=\lambda_i(E,J,L,m)$ are also all finite. Similarly, it can be easily found that
\begin{eqnarray}
V_{eff}(r=1)\mid_{\epsilon\rightarrow0}=-E^2<0 .
\end{eqnarray}
Along with \eqref{v0}, we can reach that within the interval $0 < r < 1$, there  does exist a root for $V_{eff}(r)=0$, which means that an ingoing spinning particle will turn back within the interval $0 < r < 1 $, making it possible for the particle to collide with another ingoing particle at $r=1$.
\section{Collision of Spinning Particles near Kerr Naked Singularity}\label{kerrs}
In this section, we will concentrate on the $E_{cm}$ of two colliding spinning particles in a Kerr spacetime containing a naked singularity to figure out whether there also can be a infinite $E_{cm}$ and discuss the possibility of collision between ingoing and outgoing particle at the place where the extremal horizon is location. The Kerr metric reads \cite{Wald}
\begin{eqnarray}
ds^2&=&-\left(\frac{\Delta-a^2\sin{\theta}^2}{\Sigma}\right)dt^2-\frac{2a\sin{\theta}^2(r^2+a^2-\Delta)}{\Sigma}dtd\phi\nonumber\\
&&+\sin{\theta}^2\left[\frac{(r^2+a^2)^2-\Delta a^2\sin{\theta}^2}{\Sigma}\right]d\phi^2+\frac{\Sigma}{\Delta}dr^2+\Sigma d\theta^2,
\end{eqnarray}
where
\begin{eqnarray}
&&\Sigma=r^2+a^2\cos{\theta}^2,\\
&&\Delta=r^2+a^2-2Mr,
\end{eqnarray}
the tetrad takes in form of \cite{Motoyuki}
\begin{eqnarray}\label{ei}
&&e_a^{(0)}=\sqrt{\frac{\Delta}{\Sigma}}\left(dt_a-a\sin{\theta}^2d\phi_a\right),\nonumber\\
&&e_a^{(1)}=\sqrt{\frac{\Sigma}{\Delta}}dr_a,\nonumber\\
&&e_a^{(2)}=\sqrt{\Sigma}d\theta_a,\nonumber\\
&&e_a^{(3)}=\frac{\sin{\theta}}{\sqrt{\Sigma}}\left[-adt_a+(r^2+a^2)d\phi_a\right].
\end{eqnarray}
For the convenience of following discussion, we define
\begin{eqnarray}
\mu^a=\frac{P^a}{m},
\end{eqnarray}
thus $\mu^a\mu_a=-1$. Then \eqref{scq} will becomes
\begin{eqnarray}\label{em}
-\frac{E}{m}=\mu^a\xi_a-\frac{1}{2m}S^{ab}\bigtriangledown_a\xi_b,
\end{eqnarray}
\begin{eqnarray}\label{lm}
\frac{L}{m}=\mu^a\phi_a-\frac{1}{2m}S^{ab}\bigtriangledown_a\phi_b.
\end{eqnarray}
By calculating the tetrad components of $S^{ab}$, \eqref{em} and \eqref{lm} are given by \cite{Motoyuki}
\begin{eqnarray}\label{elm}
\frac{E}{m}&=&\sqrt{\frac{\Delta}{\Sigma}}\mu^{(0)}+\frac{a\sin{\theta}}{\sqrt{\Sigma}}\mu^{(3)}+\frac{M(r^2-a^2\cos{\theta}^2)S^{(1)(0)}+2Mar\cos{\theta}S^{(2)(3)}}{m\Sigma^2},\nonumber\\
\frac{L}{m}&=&a\sin{\theta}^2\sqrt{\frac{\Delta}{\Sigma}}\mu^{(0)}+\frac{(r^2+a^2)\sin{\theta}}{\sqrt{\Sigma}}\mu^{(3)}\nonumber\\
&&+\frac{a\sin{\theta}^2}{\Sigma^2}\left[(r-M)\Sigma +2Mr^2\right]\frac{S^{(1)(0)}}{m}+\frac{a\sqrt{\Delta }\sin{\theta}\cos{\theta}S^{(2)(0)}}{\Sigma m}\nonumber\\
&&+\frac{\cos{\theta}}{\Sigma^2}\left[(r^2+a^2)^2-a^2\Delta \sin{\theta}^2\right]\frac{S^{(2)(3)}}{m}+\frac{r\sqrt{\Delta}\sin{\theta}}{m\Sigma}S^{(1)(3)}.
\end{eqnarray}
Here we also fix $\theta=\frac{\pi}{2}$, hence $\mu^{(2)}=0$. We first introduce a special Spin vector $s^{(a)}$ which reads
\begin{eqnarray}
s^{(a)}=-\frac{1}{2m}\varepsilon^{(a)}_{~~~(b)(c)(d)}\mu^{(b)}S^{(c)(d)},
\end{eqnarray}
equivalently ,
\begin{eqnarray}
S^{(a)(b)}=m\varepsilon^{(a)(b)}_{~~~~~~(c)(d)}\mu^{(c)}s^{(d)},
\end{eqnarray}
where $\varepsilon_{(a)(b)(c)(d)}$ is a totally antisymmetric tensor with component $\varepsilon_{(1)(2)(3)(4)}=1$. As is argued in \cite{Motoyuki}, we can set the only non-zero component of $s^{(a)}$ as
\begin{eqnarray}
s^{(2)}=-s ,
\end{eqnarray}
where $s$ implies both the magnitude and direction of the particle's spin. The particle spin is parallel to the black hole spin for $s>0$, while it is antiparallel for $s<0$. Then we can get these non-vanishing components of the spin angular momentum as below
\begin{eqnarray}\label{su}
&&S^{(0)(1)}=-ms\mu^{(3)},\nonumber\\
&&S^{(0)(3)}=ms\mu^{(1)},\nonumber\\
&&S^{(1)(3)}=ms\mu^{(0)}.
\end{eqnarray}
Next, we define
\bea
e=\frac{E}{m}\quad \text{and }\quad l=\frac{L}{m},
\eea
as the energy per unit mass and angular momentum per unit mass. By virtue of \eqref{su},  \eqref{elm} can be rewritten as
\begin{eqnarray}\label{eu}
e=\frac{\sqrt{\Delta}}{r}\mu^{(0)}+\frac{ar+Ms}{r^2}\mu^{(3)},
\end{eqnarray}
\begin{eqnarray}\label{lu}
l=\frac{\sqrt{\Delta}}{r}\left(a+s\right)\mu^{(0)}+\left[\frac{r^2+a^2}{r}+\frac{as}{r^2}(r+M)\right]\mu^{(3)}.
\end{eqnarray}
Combining  \eqref{eu}\eqref{lu} with the normalization condition
\begin{eqnarray}
-(\mu^{(0)})^2+(\mu^{(1)})^2+(\mu^{(3)})^2=-1,
\end{eqnarray}
we can obtain that
\begin{eqnarray}
&&\mu^{(0)}=\frac{N}{\sqrt{\Delta}(r^3-M s^2)},\label{u0}\\
&&\mu^{(1)}=\pm\frac{\sqrt{N^2-K\Delta}}{\sqrt{\Delta}(r^3-Ms^2)},\label{u1}\\
&&\mu^{(3)}=-\frac{r^2 (a e+e s-l)}{r^3-M s^2},\label{u3}
\end{eqnarray}
where
\begin{eqnarray}
N&=&r [a^2 e r+a (e s (M+r)-l r)+e r^3-l M s],\\
K&=&2Mr^3s^2-l^2r^4-r^6-M^2s^4+2alr^4e\nonumber\\
&&+2lr^4se-a^2r^4e^2-2ar^4se^2-r^4s^2e^2,
\end{eqnarray}
thus apparently, $N$ , $K$ are both finite. What's more, required by the condition that $\mu^{(a)}$ is future-directed\cite{myg}, $N > 0$  .
Here , according to \eqref{ecm}, we can define that
\begin{eqnarray}
E_{eff}=-g_{ab}\mu^a\mu^b=-\mu^{(a)}\mu_{(a)},
\end{eqnarray}
after taking \eqref{u0} \eqref{u1}and \eqref{u3} into it, also by setting $M=1$, we can get that
\begin{eqnarray}\label{efkr}
E_{eff}&=&\frac{N_1N_2\pm\sqrt{N_1^2-K_1\Delta}\sqrt{N_2^2-K_2\Delta}}{(r^3-s_1^2)(r^3-s_2^2)\Delta}\nonumber\\
&&-\frac{r^4(ae_1-l_1+e_1s_1)(ae_2-l_2+e_2s_2)}{(r^3-s_1^2)(r^3-s_2^2)},
\end{eqnarray}
Near the horizon $\Delta=0 $, we can re-express $E_{eff}$ as
\begin{eqnarray}
E_{eff}=\frac{N_1N_2\pm N_1N_2}{(r^3-s_1^2)(r^3-s_2^2)\Delta}+\gamma_{0\pm}+\gamma_{1\pm}\Delta .
\end{eqnarray}
As is discussed in\cite{Wald1972}\cite{myg} and \cite{Moller}, the relation between $s$ and $M$ is
\begin{eqnarray}
s=\frac{S}{m}<<M=1,
\end{eqnarray}
where $S$ is the particle's total spin angular momentum,  and we know that $r\approx M= 1$, thus
\begin{eqnarray}
r^3-s^2 > 0.
\end{eqnarray}
So it is safe to assert that both $\gamma_{0\pm}$ and $\gamma_{1\pm}$ are finite.

Now let's consider the naked singularity, we assume that $a$ is slightly larger than 1, i.e. we can set that $a=1+\epsilon$  as we did before, so near $r=1$ where the original extremal horizon is located, $ \Delta $ can be rewritten as\cite{naked}
\begin{eqnarray}
\Delta\mid_{r=1}=2\epsilon .
\end{eqnarray}
So we choose the $Plus$ sign in \eqref{efkr}, we will get that
\begin{eqnarray}\label{effkr}
E_{eff}=\frac{N_1N_2}{(r^3-s_1^2)(r^3-s_2^2)\epsilon}+\text{finite terms} ,
\end{eqnarray}
thus $E_{eff}$ can be arbitrarily high as $\epsilon\rightarrow 0$ .

What we should do next is to find the condition under which an ingoing spinning particle will turn back within $0 < r < 1 $, so that the $E_{eff}$ will behave as \eqref{effkr}. To define Effective Potential  $V_{eff}$  as above, we need to figure out the radial component of $\mu^a$, which is
\begin{eqnarray}
\mu^r=\mu_a\left(\frac{\partial}{\partial r}\right)^a=\mu_{(a)}\left(\frac{\partial}{\partial r}\right)^a=\left[\Sigma_{i=0}^3\mu^{(i)}e_a^{(i)}\right]\left(\frac{\partial}{\partial r}\right)^a.
\end{eqnarray}
Based on \eqref{ei}\eqref{u0}\eqref{u1}\eqref{u3} , we can obtain that
\begin{eqnarray}
\mu^r=\pm\frac{\sqrt{N^2-K\Delta}\sqrt{\Sigma}}{\Delta (r^3-s^2)},
\end{eqnarray}
then we can define  $V_{eff}$ as
\begin{eqnarray}\label{vu1}
V_{eff}=-(\mu^r)^2=-\frac{(N^2-K\Delta)\Sigma}{\Delta^2 (r^3-s^2)^2}.
\end{eqnarray}
Since  $\theta=\frac{\pi}{2}$, thus $\Sigma=r^2 $, then we can easily find that
\begin{eqnarray}
V_{eff}\mid_{r=0}=0.
\end{eqnarray}
After taking $a=1+\epsilon$, $r=1$ into \eqref{vu1}, $V_{eff}$ becomes
\begin{eqnarray}
V_{eff}\mid_{r=1}=-\frac{(1+s)^2(l-2e)^2}{4\epsilon^2(1-s^2)^2},
\end{eqnarray}
where in the numerator, we have set $\epsilon=0$ . From above, we can see that when $l=2e$, $V_{eff}\mid_{r=1}=0$, which is consistent with the result in \cite{myg}; when $l\neq 2e $,
\bean
V_{eff}\mid_{r=1}\rightarrow -\infty \quad \text{as} \quad \epsilon\rightarrow 0.\nonumber
\eean
Since merely $V_{eff}\mid_{r=0}$ and $V_{eff}\mid_{r=1}=0 $ cannot guarantee a solution within $ 0 < r < 1 $ for equation $V_{eff}=0 $, we need to further discussion the specific form of $V_{eff}$ which can be written explicitly as
\begin{eqnarray}\label{vkr}
V_{eff}=\frac{r^2[(1-e^2)r^8-2r^7+\Sigma_{i=1}^6\omega_i r^i+s^4]}{\Delta^2 (r^3-s^2)^2},
\end{eqnarray}
where all $\omega_i$ are finite. In \eqref{vkr}, the numerator behaves as
\begin{eqnarray}
[(1-e^2)r^8-2r^7+\Sigma_{i=1}^6\omega_i r^i+s^4]\mid_{r\rightarrow0}=s^4>0,
\end{eqnarray}
hence, we can conclude that near $ r=0 $ there must exist an small interval $(0,\delta)$ ( usually $\delta<<1$ ) within which
\begin{eqnarray}
(1-e^2)r^8-2r^7+\Sigma_{i=1}^6\omega_i r^i+s^4>0.
\end{eqnarray}
Namely , $V_{eff}>0$ within this interval. So, combining with $V_{eff}\mid_{r=1}<0 $, we can approach that equation $V_{eff}=0$ has a root within $(\delta , 1)$. Therefore, an ingoing  particle can turn back within the interval $0 < r < 1 $, meaning that the collision leading to a infinite $E_{cm}$ is possible to happen.
\section*{Conclusion}
In an extremal Reissner-Nordstrom black hole background($Q=M$), if the black hole become slightly over-charged($Q=M+\epsilon$), the horizon located at $r=M$ will vanish abruptly, the singularity at $r=0$ becomes uncovered, which is just a naked singularity. Similarly, when an extremal Kerr black hole ($a=M$) gets slightly over-spun($a=M+\epsilon$), the horizon will also vanish,  hence there exists a naked singularity too. The discussion in this paper is mainly focus on such kind of spacetimes that admit naked singularities.

In this paper, it was proved that an ingoing spinning  particle released from infinity towards the Reissner-Nordstrom or Kerr naked singularities can certainly turn back after it pass $r = M $ where the original extremal event horizon is located before it reach the naked singularities. Consequently, it is possible for the turning-back spinning particle to collide with another ingoing particle. If the spacetime is slightly over-spun or over-charged, i.e. $\epsilon$ is arbitrarily small, the $E_{cm} $ can be arbitrarily large, so a spinning particle can be accelerated to arbitrarily high energy by a RN or Kerr naked singularity.

\section{Acknowledgement}
The work was supported by NSFC Grants No. 11375026 and 11235003.


\begin{thebibliography}{99}
\bibitem{BSW}M.Ba\~nados, J.Silk and S.M.West, Phys. Rev. Lett. {\bf 103,} 111102(2009).
\bibitem{272}O. B. Zaslavskii, Phys. Rev. D {\bf82,} 083004 (2010).
\bibitem{273}O. B. Zaslavskii, Classical Quantum Gravity {\bf28,} 105010 (2011).
\bibitem{274}S. W. Wei, Y.X. Liu, H. Guo, and Chun-E Fu, Phys. Rev. D {\bf82,} 103005 (2010).
\bibitem{275}S. W. Wei, Y. X. Liu, H. Guo, and Chun-E Fu, J. High Energy Phys. 12 (2010) 066.
\bibitem{276}M. Kimura, K. Nakao, and H. Tagoshi, Phys. Rev. D {\bf83,} 044013 (2011).
\bibitem{277}T. Harada and M. Kimura, Phys. Rev. D {\bf83,} 084041 (2011).
\bibitem{278}Changchun Zhong, Sijie Gao,JETP Letters, 2011, Vol. 94, No. 8, pp. 589-592.
\bibitem{1025}A. Galajinsky,  Phys. Rev. D {\bf88,}  027505(2013).
\bibitem{4D}Sijie Gao, Changchun Zhong, Phys.Rev. D {\bf84,} 044006 (2011).
\bibitem{Wald1972}R.M.Wald, Phys.Rev. D {\bf6,} 406(1972).
\bibitem{myg}M.-Y. Guo, and S.-J. Gao, Phys. Rev. D {\bf93,} 084025(2016).
\bibitem{ks}
  J.~An, J.~Peng, Y.~Liu and X.~H.~Feng,
  arXiv:1710.08630.
  \bibitem{zs}
  O.~B.~Zaslavskii,
  EPL {\bf 114}, no. 3, 30003 (2016).
\bibitem{Wald}R.M.Wald, General Relativity (The University of Chicago Press, Chicago, 1984).
\bibitem{Motoyuki}Motoyuki Saijo, Kei-ichi Maeda, Masaru Shibata and YasushiMino, Phys.Rev. D {\bf58,} 064005 (1998).
\bibitem{Moller}C. Moller, Commun. Dublin Inst.Advan.Stud. A5, 1 (1949).
\bibitem{naked}M. Patil and P. S. Joshi, Classical Quantum Gravity {\bf28,} 235012 (2011).
\bibitem{naked1}Mandar Patil, Pankaj S. Joshi, Phys. Rev. D {\bf85,} 104014 (2012).
\bibitem{RNnaked}Mandar Patil, Pankaj S. Joshi, Masashi Kimura, Ken-ichi Nakao,Phys. Rev. D {\bf86,} 084023 (2012).
\bibitem{naked2}Mandar Patil, Pankaj S. Joshi, arXiv:1106.5402.
\bibitem{s1}M.Mathisson, Acta Phys. Pol. 6, 163 (1937).
\bibitem{s2}A.Papapetrou, Proc. Roy. Soc.(London) A 209, 248 (1951).
\bibitem{s3}W.G.Dixon, Proc. Roy. Soc.(London) 314, 499(1970).
\bibitem{z1}E.Hackmann, Cl.Lammerzahl, Yu.N.Obukhov, D.Puetzfeld, I.Schaffer,Phys. Rev. {\bf90,} 064035(2014).
\bibitem{z2}P.I.Jefremov, O.Y.Tsupko, G.S.Bisnovatyi-Kogan, Phys.Rev. D {\bf91,} 124030 (2015).
\bibitem{yuxiao}Yu-Peng Zhang, Bao-Min Gu, Shao-Wen Wei, Jie Yang, Yu-Xiao Liu , Phys. Rev. D {\bf 94,} 124017 (2016).

\end{thebibliography}
\end{document}